\documentclass[conference,9pt]{IEEEtran}
\IEEEoverridecommandlockouts
\usepackage{cite}
\usepackage{amsmath,amssymb,amsfonts}
\usepackage{graphicx}
\usepackage{subcaption}
\usepackage{textcomp}
\usepackage{amsmath}
\usepackage{booktabs} 
\usepackage{multirow}
\usepackage[linesnumbered,ruled,vlined]{algorithm2e}

\usepackage{hyperref}       
\usepackage{url}            
\usepackage{booktabs}       
\usepackage{amsfonts}       
\usepackage{nicefrac}       
\usepackage{microtype}      
\usepackage{xcolor}         
\usepackage{amsmath}
\usepackage{tikz}

\usepackage{orcidlink}
\usepackage{comment}
\usepackage{amssymb}
\usepackage{braket}
\usepackage{xcolor}
\usepackage{adjustbox}

\def\BibTeX{{\rm B\kern-.05em{\sc i\kern-.025em b}\kern-.08em
    T\kern-.1667em\lower.7ex\hbox{E}\kern-.125emX}}
\begin{document}

\title{CircuitHunt: Automated Quantum Circuit Screening for Superior Credit-Card Fraud Detection}

\author{\IEEEauthorblockN{Nouhaila Innan\textsuperscript{1,2}, 
Akshat Singh\textsuperscript{3},  Muhammad Shafique\textsuperscript{1,2}
\IEEEauthorblockA{
\textsuperscript{1}eBRAIN Lab, Division of Engineering, New York University Abu Dhabi (NYUAD), Abu Dhabi, UAE\\
\textsuperscript{2}Center for Quantum and Topological Systems (CQTS), NYUAD Research Institute, NYUAD, Abu Dhabi, UAE\\
\textsuperscript{3}New York University Tandon School of Engineering, New York, US\\
$\{$nouhaila.innan, akshat.singh, muhammad.shafique$\}$@nyu.edu\\
}
}}

\maketitle
\begin{abstract}
Designing effective quantum models for real-world tasks remains a key challenge within Quantum Machine Learning (QML), particularly in applications such as credit card fraud detection, where extreme class imbalance and evolving attack patterns demand both accuracy and adaptability. Most existing approaches rely on either manually designed or randomly initialized circuits, leading to high failure rates and limited scalability. In this work, we introduce CircuitHunt, a fully automated quantum circuit screening framework that streamlines the discovery of high-performing models. CircuitHunt filters circuits from the KetGPT dataset using qubit and parameter constraints, embeds each candidate into a standardized hybrid QNN, and performs rapid training with checkpointing based on macro-F1 scores to discard weak performers early. The top-ranked circuit is then fully trained, achieving 97\% test accuracy and a high macro-F1 score on a challenging fraud detection benchmark. By combining budget-aware pruning, empirical evaluation, and end-to-end automation, CircuitHunt reduces architecture search time from days to hours while maintaining performance. It thus provides a scalable and task-driven tool for QML deployment in critical financial applications.
\end{abstract}


\section{Introduction}

Credit card fraud detection remains one of the most mission-critical yet technically challenging domains in financial technology \cite{hafez2025systematic}. The key problem we address in this paper is: \textbf{How can quantum machine learning (QML) models be efficiently and automatically designed to perform well on real-world classification tasks characterized by extreme class imbalance, such as fraud detection?} 
This challenge extends beyond model selection; it is fundamentally a pipeline design issue. In such datasets, fraudulent transactions often constitute less than 1\% of total activity, making standard supervised learning pipelines susceptible to bias and underperformance \cite{berkmans2025anomaly,grossi2022mixed,tudisco2024evaluating}.

While oversampling techniques such as the synthetic minority over-sampling technique (SMOTE) can help rebalance the input distribution \cite{chawla2002smote}, they only address data imbalance. They do not mitigate the architectural sensitivity of quantum learning models. Moreover, fraud behaviors continuously evolve in adversarial and non-stationary ways, demanding detection systems that are accurate, adaptable, and resource-aware. These real-world constraints position fraud detection as a natural testbed for emerging learning paradigms, including QML.

Within the QML landscape \cite{biamonte2017quantum,bowles2024better}, hybrid quantum neural networks (QNNs), which combine classical preprocessing with parameterized quantum circuits (PQCs), have shown early promise \cite{abbas2021power,innan2025next}. These models can potentially leverage the unique representational capacity of quantum states, enabling richer mappings even with few qubits \cite{kashif2024computational}. However, a persistent bottleneck remains: the design and selection of high-performing quantum circuits \cite{ahmed2025quantum}. Unlike classical neural networks, which benefit from well-established architectures and scalable neural architecture search (NAS), the PQC landscape remains fragmented, with no clear guidelines for depth, topology, or parameterization. Performance depends heavily on circuit structure, and most existing research either (i) hand-designs a small set of circuits using domain knowledge or (ii) generates random PQCs without considering trainability or task suitability.

Several recent studies have investigated QML models for classification \cite{gong2024quantum,innan2023enhancing,wu2023more,innan2024variational, innan2024lep,siddiqui2024quantum,dave2025sentiqnf}. Some propose hand-crafted circuits using heuristic encoding and entanglement patterns; others explore training small ensembles of randomly initialized circuits. While these approaches demonstrate feasibility, they suffer from several limitations:
\begin{itemize}
 \item \textbf{Lack of diversity:} Hand-designed circuits often reflect limited structural exploration.
 \item \textbf{Inefficient search:} Random circuit trials require extensive simulation and often result in poorly converging or untrainable models.
 \item \textbf{Scalability issues:} As dataset complexity or qubit budgets grow, manual design quickly becomes infeasible.
\end{itemize}

Recent datasets like KetGPT offer a promising shift, providing 1,000 structurally valid quantum circuits generated from transformer models trained on real OpenQASM programs \cite{apak2024ketgpt}. These circuits span a wide range of qubit counts and gate complexities, capturing realistic design patterns beyond what manual creation can achieve. However, no existing framework efficiently integrates such datasets into an automated, task-driven architecture screening process.
\subsection{Motivational Case Study: Why We Need Automated Circuit Screening}
To quantify the architectural bottleneck in quantum model design, we conducted a preliminary analysis using 3 manually crafted circuits and more than 10 circuits generated by KetGPT without any filtering or screening. Integrated into a fixed hybrid model and evaluated using macro-F1 score on a binary fraud detection task, only 33\% of the manually designed or randomly initialized circuits produced meaningful gradients or converged to acceptable performance (macro-F1 score $>$ 0.5). In contrast, over 85\% of the KetGPT-generated circuits were trainable and executable, though performance varied widely, with macro-F1 scores ranging from 0.325 to 0.686.

These results reveal two key insights:
\begin{itemize}
\item Manual or random circuit selection often results in high failure rates and inefficient resource usage.

\item  Even among valid circuits, empirical evaluation is necessary to identify high-performing candidates.
\end{itemize}
Automated architecture discovery, guided by simulator/hardware constraints and validation-driven screening, is therefore essential for scalable, robust QML deployment.
\subsection{Scientific Challenges and Novel Contributions}
In light of the above, we target the following scientific challenges:
\begin{itemize}
    \item 
\textbf{C1:} Navigating a large search space of real quantum circuits under practical constraints such as qubit count, parameter budget, and simulation cost.

  \item \textbf{C2:} Identifying circuits that are not only structurally valid but also empirically high-performing across different learning conditions.

  \item \textbf{C3:} Enabling automation that reduces human effort and time-to-deployment from days/weeks to hours without compromising accuracy.
\end{itemize}
To address these challenges, we propose CircuitHunt, an automated quantum circuit screening framework designed to identify high-performing circuits from a structured dataset, using empirical feedback and budget-aware filtering.

\textbf{Our key contributions are:}

\begin{itemize}
    \item \textbf{Constraint-Aware Circuit Filtering:}
We design a filtering pipeline that excludes circuits exceeding qubit or parameter limits, lacking trainable gates, or failing execution, ensuring only viable candidates are considered.

\item \textbf{Macro-F1 Score Based Selection:}
We embed each circuit in a standardized hybrid QNN and perform brief training runs. Validation macro-F1 score is used to efficiently eliminate underperforming architectures.

\item \textbf{Modular End-to-End Pipeline:}
We build an end-to-end framework, ``CircuitHunt'', that integrates preprocessing, quantum module selection, residual architecture with learnable skip connections, and post-processing into a reproducible workflow.

\item \textbf{Strong Performance and Ablation Analysis:}
We achieve 97\% test accuracy while maintaining a high macro-F1 score, and our ablation studies highlight the contributions of skip connections and constraint-based filtering.
\end{itemize}
\section{Related Work}
\subsection{QML for Fraud Detection}
QML has demonstrated promising potential and a variety of proof-of-concept studies across multiple sectors \cite{innan2024quantum33,siddiqui2024quantum,el2024quantum,innan2025optimizing,innan2025qnn,pathak2024resource,dutta2025quiet}, including finance \cite{dutta2024qadqn,choudhary2025hqnn,doosti2024brief}. In particular, fraud detection—our primary focus—poses unique challenges such as high-dimensional, imbalanced data and stringent performance requirements. Recent studies have validated the effectiveness of hybrid QML architectures, including QNNs, in capturing complex fraud patterns \cite{Innan_2024,innan2024financial1,alami2024comparative, tudisco2024evaluating, ubale2025toward,innan2024qfnn,sawaika2025privacy}. However, despite these advances, most existing approaches continue to rely on manually designed circuit architectures, often based on fixed ansatz and limited architectural exploration. Crucially, none of the current works integrate automated architecture discovery or circuit search strategies tailored to specific task requirements. As datasets become more diverse and application demands become more rigorous, manual circuit tuning becomes increasingly impractical. There is a growing need for scalable, automated methods capable of generating, evaluating, and selecting circuit architectures based on application-level performance objectives. This highlights the importance of QAS frameworks that not only respect hardware constraints but also optimize architectures for real-world deployment, representing a necessary step forward in building robust, task-aware QML systems.

\subsection{Quantum Architecture Search}
Quantum architecture search (QAS) has emerged as a promising direction to automate the design of PQCs \cite{zhang2022differentiable,dutta2025qas}, aiming to replace manual circuit crafting with data- or task-driven optimization strategies. Existing QAS approaches primarily focus on reducing circuit depth, improving trainability under noise, and navigating barren plateaus through either search-based heuristics or reinforcement learning frameworks.

Several notable works have demonstrated the potential of QAS to optimize variational ansatz for specific learning tasks or physical simulations. One line of research proposed frameworks that balanced expressivity and noise accumulation, demonstrating improved robustness over manually designed VQAs in tasks such as classification and quantum chemistry \cite{du2022quantum}. Comprehensive surveys outlined the complexity and resource demands involved in designing effective PQCs tailored to both hardware constraints and task requirements \cite{martyniuk2024quantum}.

Other studies applied reinforcement learning to construct circuits for target states or operations. Some methods used probabilistic policy reuse integrated with deep Q-learning to synthesize circuits under diverse noise conditions, improving generalization through continual learning \cite{ye2021quantum}. Others trained agents using actor-critic and proximal policy optimization algorithms to generate efficient gate sequences without relying on physics-based priors, demonstrating effectiveness in multi-qubit state preparation tasks such as GHZ or Bell states \cite{kuo2021quantum}. Reinforcement learning environments for quantum circuit design have also been formalized as Markov decision processes \cite{altmann2024challenges}, enabling the development and benchmarking of policies capable of controlling universal, continuously parameterized gate sets.

Despite these advances, most QAS studies remain largely decoupled from real-world application requirements. Architectures are often evaluated on standard benchmark datasets with idealized or simulated conditions, with limited emphasis on application-specific performance metrics or domain-relevant constraints (e.g., imbalanced data in fraud detection, or interpretability in finance). Additionally, the use of limited datasets can create overfitting to specific tasks while overlooking the generalization capacity of the circuit.

Given the increasing availability of diverse quantum and classical datasets, architecture discovery must go beyond theoretical search efficiency and incorporate empirical screening based on target applications. This includes using broader datasets, realistic training pipelines, and domain-specific evaluation criteria (e.g., macro-F1 score for imbalanced classification). Our work addresses these gaps by introducing a QAS framework that combines hardware-aware filtering with application-driven validation, ensuring that selected circuits are not only feasible but also effective in practical deployment settings.

\section{CircuitHunt}
\begin{figure}[htbp]
    \centering
    \includegraphics[width=\linewidth]{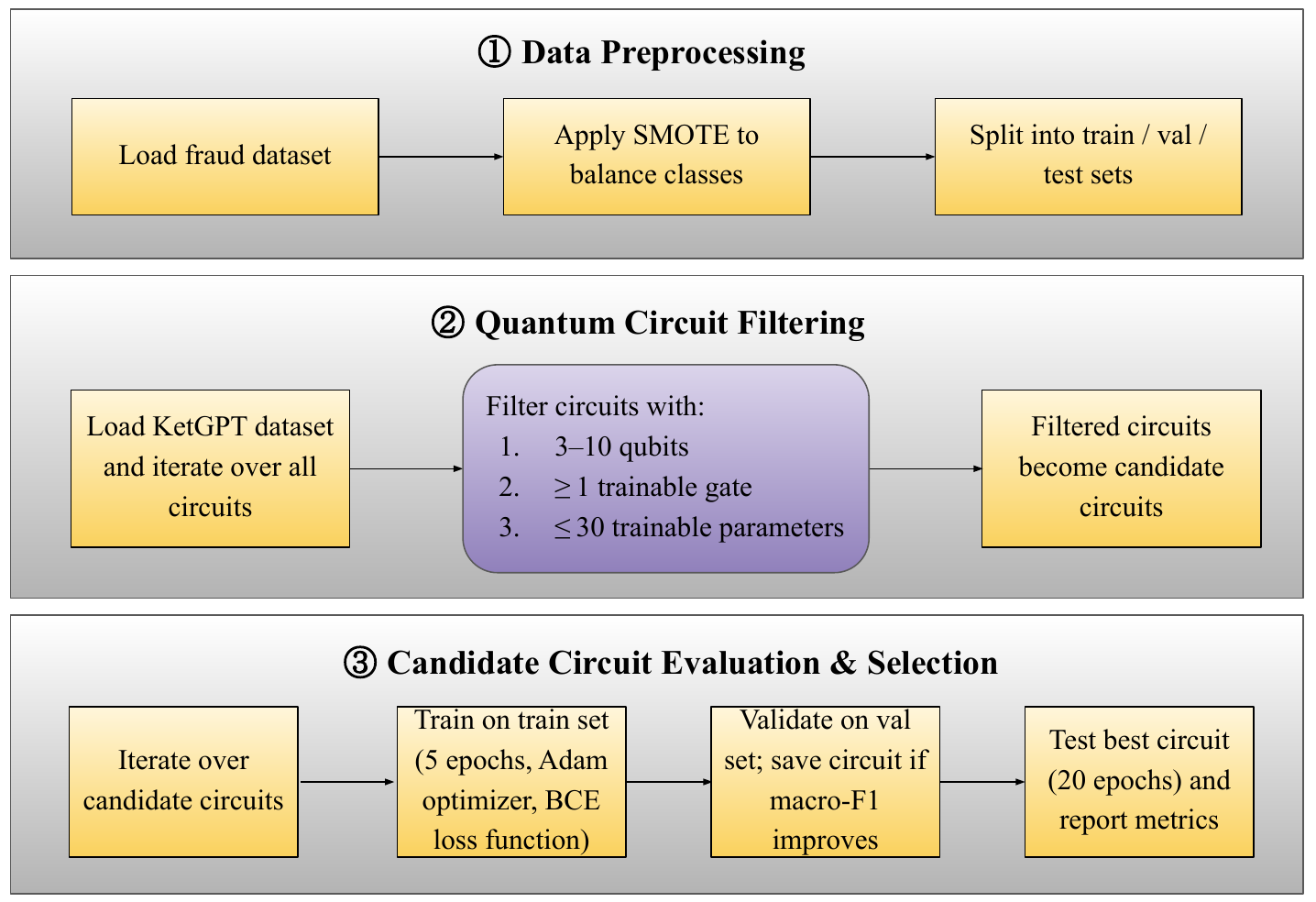}
    \vspace{-0.5cm}
    \caption{ \footnotesize End-to-end pipeline for hybrid quantum-classical circuit evaluation. 
    (1) The credit card fraud dataset is preprocessed with SMOTE-based balancing, scaling, and split into training, validation, and test sets.
    (2) Candidate quantum circuits are filtered from the KetGPT dataset by qubit count (3–10), presence of trainable gates, and parameter budget ($\leq$ 30).
    (3) Each candidate circuit is evaluated using a hybrid model trained for 5 epochs. If validation macro-F1 score improves, the circuit is checkpointed. After all circuits are evaluated, the circuit selected by CircuitHunt is tested on the held-out test set.}
    \label{fig:pipeline}
\end{figure}
This section presents our proposed methodology in detail. As illustrated in Fig.~\ref{fig:pipeline}, CircuitHunt proceeds in three main stages: data preprocessing, quantum circuit filtering, and candidate circuit evaluation and selection (see Algorithm \ref{CircuitHunt}).

\subsection{Data Preprocessing}
The preprocessing stage begins by separating features from the target labels. Given the highly imbalanced nature of the dataset (1\% positive class), we apply SMOTE to increase the representation of the minority class and address the class imbalance. To construct a balanced and computationally manageable dataset, an equal number of samples are then randomly selected from each class.
The dataset's features are already transformed using principal component analysis (PCA), which ensures decorrelation and dimensionality reduction. As such, no additional feature engineering or transformation is required. All feature values are then scaled using Min-Max normalization to a bounded interval, ensuring consistent input magnitudes and facilitating compatibility with quantum processing stages. The normalized dataset is subsequently partitioned into training, validation, and test sets using stratified sampling, which preserves class proportions and supports reliable model evaluation.

\subsection{Quantum Circuit Filtering}

The second stage of the \textit{CircuitHunt} framework involves selecting suitable quantum circuits from a structured dataset to serve as candidates for hybrid learning. Unlike randomly generated quantum circuits, which often lack coherence or trainability, we use the KetGPT dataset, which includes circuits with qubit counts ranging from 2 to 117 and gate counts between 6 and 903~\cite{pennylaneKetGPT}, offering a diverse search space to \textit{hunt for} promising candidates that can empirically improve performance on our target problem, credit card fraud detection.

Let $\mathcal{C} = \{C_1, C_2, \dots, C_N\}$ denote the initial set of quantum circuits, where each circuit $C_i$ is defined over $n_i$ qubits and consists of a sequence of quantum operations $\{G_{i1}, G_{i2}, \dots, G_{im_i}\}$. The objective is to construct a filtered subset $\mathcal{C}' \subset \mathcal{C}$ such that each retained circuit satisfies a set of hardware and learning constraints:

\begin{itemize}
    \item \textbf{Qubit Budget Constraint:} To ensure compatibility with current quantum hardware and simulation resources, we retain circuits with: $n_{\min} \leq n_i \leq n_{\max}$,
where $n_{\min} = 3$, $n_{\max} = 10$.
    \item \textbf{Parameterized Gate Constraint:} A circuit must contain at least one trainable gate. Let $\mathcal{G}_\text{trainable} = \{\mathrm{R_Y}(\theta), \mathrm{R_Z}(\theta), U_2(\phi, \lambda)\}$ denote the set of parameterized gates. We require:
     \begin{equation}
        \exists \, G_{ij} \in C_i \text{ such that } G_{ij} \in \mathcal{G}_\text{trainable}.
     \end{equation}

    \item \textbf{Parameter Count Constraint:} The total number of trainable parameters in a circuit must not exceed a threshold:
    \begin{equation}
        P_i = \sum_{j=1}^{m_i} \delta(G_{ij}), \quad \text{with } P_i \leq P_{\max},
   \end{equation}
    where $\delta(G_{ij})$ denotes the number of parameters for gate $G_{ij}$. 

    \item \textbf{Execution Validation:} Each filtered candidate is embedded in a QNode and simulated using zero-initialized parameters. The circuit is retained only if the expectation value is finite and within bounds:
   \begin{equation}
        f_i(\boldsymbol{\theta}) = \left\langle \psi_i(\boldsymbol{\theta}) \middle| Z_0 \middle| \psi_i(\boldsymbol{\theta}) \right\rangle \in \mathbb{R}, \quad \text{and} \quad |f_i(\boldsymbol{\theta})| \leq 1.
   \end{equation}
\end{itemize}

Circuits failing any of the above criteria are excluded. The resulting subset $\mathcal{C}'$ contains expressive, parameter-efficient, and executable circuits grouped by qubit count. These serve as input to the subsequent training phase, where their empirical performance is evaluated on the fraud detection task.

\begin{figure}[htbp]
    \centering
    \includegraphics[width=\linewidth]{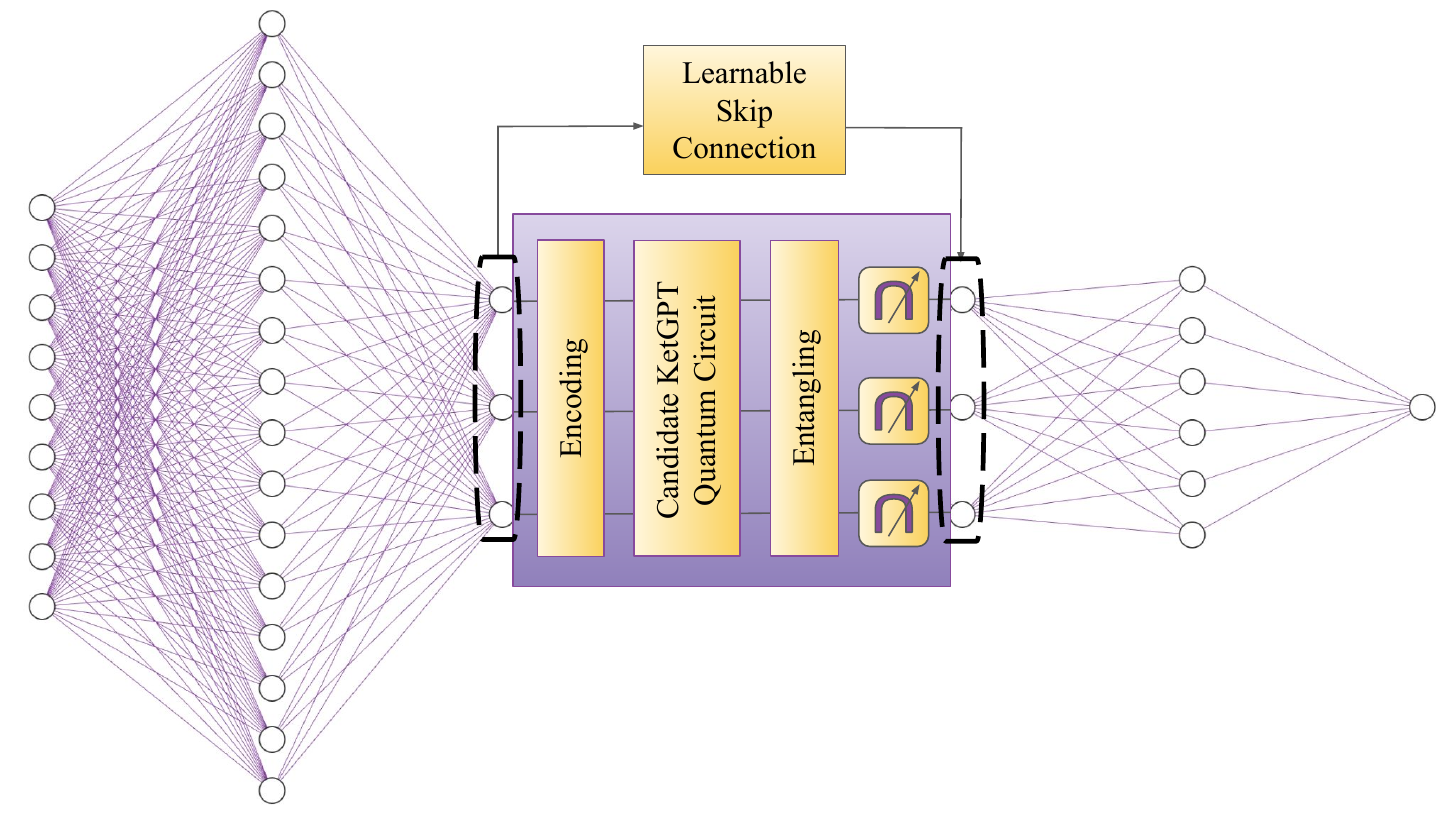}
    \vspace{-0.6cm}
    \caption{\footnotesize Representative architecture of the hybrid quantum-classical model used to evaluate candidate circuits. The model consists of a pre-quantum classical neural network that maps $n$ input features to $q$ outputs, a quantum module using KetGPT circuits with encoding and entanglement layers, a learnable residual skip connection from classical to quantum output, and a post-quantum neural network outputting a raw logit for binary classification. 
    }
    \label{fig:model}
\end{figure}
\subsection{Candidate Circuit Evaluation \& Selection}

The third stage concerns the systematic evaluation and selection of candidate quantum circuits. This phase involves embedding each filtered circuit into a unified hybrid architecture, followed by performance benchmarking using a validation-based criterion.

To ensure fair comparison, all candidate circuits are integrated into a consistent hybrid quantum-classical model composed of the following components as shown in Fig. \ref{fig:model}:

\begin{itemize}
    \item \textbf{Pre-quantum classical network:} A feedforward neural network that transforms the normalized classical input $\mathbf{x} \in \mathbb{R}^{f}$, into a latent representation in $\mathbb{R}^{q}$, matching the qubit width of the candidate circuit. This module comprises a hidden layer of dimension  and an output layer of size $q$.

    \item \textbf{Quantum circuit module:} The output of the classical encoder is encoded into a quantum state using parameterized rotation gates (e.g., $R_X$). The encoded input is then processed by a candidate KetGPT circuit, followed by entangling operations (e.g., linear CNOT chains). The expectation values of Pauli-$Z$ observables at each qubit yield a quantum-processed vector in $\mathbb{R}^{q}$.

    \item \textbf{Learnable skip connection:} To preserve gradient flow and enhance expressivity, a learnable residual connection is introduced. The final quantum output is computed as:
    \begin{equation}        
        \mathbf{z}_{\text{res}} = \mathbf{z}_{\text{quantum}} + \alpha \cdot \mathbf{z}_{\text{classical}},
    \end{equation}
    where $\alpha \in \mathbb{R}$ is a trainable scalar parameter.

    \item \textbf{Post-quantum classical network:} The residual-enhanced vector $\mathbf{z}_{\text{res}} \in \mathbb{R}^{q}$ is passed through a final multilayer perceptron comprising a hidden layer of size $h_2 = 16$ and an output layer mapping to $\mathbb{R}$ for binary fraud classification.
\end{itemize}

 Let $\mathcal{C}' = \{C_1, C_2, \dots, C_K\}$ denote the filtered set of candidate circuits obtained from the previous stage. Each circuit $C_i \in \mathcal{C}'$ is trained independently using the hybrid architecture for a fixed number of epochs $T_{\text{short}}$ to efficiently approximate its learning capacity.

During training, we monitor the macro-averaged F1 score on the validation set:
\begin{equation}
    \text{F1}_{\text{macro}} = \frac{1}{2} \left( \text{F1}_{\text{fraud}} + \text{F1}_{\text{non-fraud}} \right),
    \end{equation}

where $\text{F1}_{\text{fraud}}$ and $\text{F1}_{\text{non-fraud}}$ denote the per-class F1 scores. After training all circuits, we select the optimal circuit $C^{*} \in \mathcal{C}'$ as:
\begin{equation}
    C^{*} = \arg\max_{C_i \in \mathcal{C}'} \, \text{F1}_{\text{val}}^{(i)}.
    \end{equation}

The selected circuit $C^{*}$ is then re-trained from scratch for an extended number of epochs $T_{\text{full}}$ using the full training dataset. Final performance is evaluated on the held-out test set, and the corresponding metrics are reported for analysis.
\begin{algorithm}[htpb]
\caption{CircuitHunt}
\label{CircuitHunt}
\footnotesize
\KwIn{Raw dataset $D$ with features and labels, KetGPT Circuits}
\KwOut{Trained HQNNs and evaluation metrics}

\SetKwFunction{TrainModel}{TrainModel}
\SetKwFunction{Evaluate}{EvaluateModel}

Load dataset $D$ and extract features $X$ and labels $y$ \\
Apply SMOTE to balance the dataset \\
Scale $X$ to $[0, \pi]$ using MinMaxScaler \\
Split into training, validation, and test sets: $\{X_{\text{train}}, X_{\text{val}}, X_{\text{test}}, y_{\text{train}}, y_{\text{val}}, y_{\text{test}}\}$ \\

Load pre-built KetGPT circuits from dataset \\
\ForEach{circuit $C$ with $n$ qubits}{
    \If{$n < 3$ or $n > 10$}{\textbf{continue}}
    Extract parameterized gates in $C$ \\
    Count parameters $P$ from gates \\
    \If{$P > 30$ or $P = 0$}{\textbf{continue}}
    \If{execution of $C$ fails or output is non-finite}{\textbf{continue}}
    Add $C$ to \texttt{trainable\_circuits}
}

Initialize variables to track best-performing model \\
\For{$n \in \{3, \dots, 10\}$}{
    Prepare datasets for $n$-qubit input size \\
    \ForEach{circuit $C$ with $n$ qubits}{
        Construct hybrid Quantum Neural Network using $C$ \\
        Initialize weights and optimizer (Adam) \\
        \TrainModel{$C$, $X_{\text{train}}, y_{\text{train}}, X_{\text{val}}, y_{\text{val}}$} \\
        Evaluate validation macro-F1 score \\
        Save checkpoint if performance improves \\
    }
}

Load best model and weights \\
\Evaluate{$X_{\text{test}}, y_{\text{test}}$}

\SetKwProg{Fn}{Function}{:}{}
\Fn{\TrainModel{$C$, $X_{\text{train}}, y_{\text{train}}, X_{\text{val}}, y_{\text{val}}$}}{
    \For{each epoch}{
        Train on mini-batches from $X_{\text{train}}, y_{\text{train}}$ using BCE loss \\
        Evaluate predictions on training and validation sets \\
        Track metrics: Accuracy, Precision, Recall, F1
    }
}

\Fn{\Evaluate{$X, y$}}{
    Predict using trained model \\
    Compute and report final test metrics
}

\end{algorithm}

\section{Results and Discussion}
\subsection{Experimental Setup}

All experiments were conducted on a MacBook Air (Model: Mac15,13) equipped with an Apple M3 chip featuring an 8-core CPU (4 performance cores and 4 efficiency cores) and 16 GB of RAM. The implementation relies on PennyLane's \texttt{default.qubit} simulator \cite{pennylane}, with quantum circuits integrated via PyTorch-compatible QNodes to ensure end-to-end differentiability.

As detailed in Table~\ref{tab:grouped-settings}, the CircuitHunt pipeline explores quantum circuits with 3 to 10 qubits and up to 30 trainable parameters. The final selected circuit, referred to as \textit{CircuitHunt-Selected}, uses 6 qubits with 9 trainable parameters and corresponds to index \#221 in the KetGPT circuit repository \cite{ketrepo}. All other architectural and training configurations remain fixed across the experiments to ensure fairness and reproducibility.

\begin{table}[ht]
\centering
\caption{Experimental Settings.}
\label{tab:grouped-settings}
\begin{adjustbox}{max width=\linewidth}
\begin{tabular}{lll}
\toprule
& \textbf{Parameter} & \textbf{Value}  \\
\midrule
\multirow{5}{*}{\rotatebox{90}{\textbf{Dataset}}} 
    & Dataset           & Credit Card Fraud \cite{data}       \\
    & Samples per Class      & 10,000                 \\
    & Number of Features     & 28                      \\
    & Classes                & (Non-Fraud = 0, Fraud = 1) \\
    & Split Ratio            & 60/10/30 (train/val/test) \\
\midrule

\multirow{3}{*}{\rotatebox{90}{\textbf{Quantum}}} 
    & Source                 & KetGPT        \\
    & Data Encoding &  Angle Encoding\\
    & Qubit Range            & $n \in [3, 10]$         \\
    & Trainable Params Max   & 30                    \\
\midrule

\multirow{5}{*}{\rotatebox{90}{\textbf{Classical}}} 
    & Pre-NN Layers          & Linear(28,64) → ReLU → Linear(64,$n$) \\
   
    & Post-NN Layers         & Linear($n$,16) → ReLU → Linear(16,1)  \\
    & Activation             & Sigmoid               \\
    & Residual Initialization Value & 0.1\\
\midrule

\multirow{6}{*}{\rotatebox{90}{\textbf{Training}}} 
    & Optimizer              & Adam                 \\
    & Learning Rate          & 0.01                  \\
    & Loss Function          & Binary Cross-Entropy with Logits    \\
    & Batch Size             & 32                   \\
    & Epochs                & 5 (init), 20 (final)    \\
&Device & \texttt{default.qubit} (PennyLane simulator) \\
&Circuit Execution Interface & PyTorch-compatible QNode \\
&Validation Metric & Macro-F1 score \\
&Model Selection & Based on best validation macro-F1 score \\    
\bottomrule
\end{tabular}
\end{adjustbox}
\end{table}

\subsection{Performance Analysis}
To evaluate the effectiveness of the CircuitHunt-selected model, we conducted a comprehensive analysis across multiple metrics, including accuracy, loss, macro-F1 score, and ROC-AUC. The results consistently highlight the model's strong generalization ability, optimization stability, and class-wise predictive reliability.
\begin{figure*}[htpb]
    \centering
    \includegraphics[width=1\linewidth]{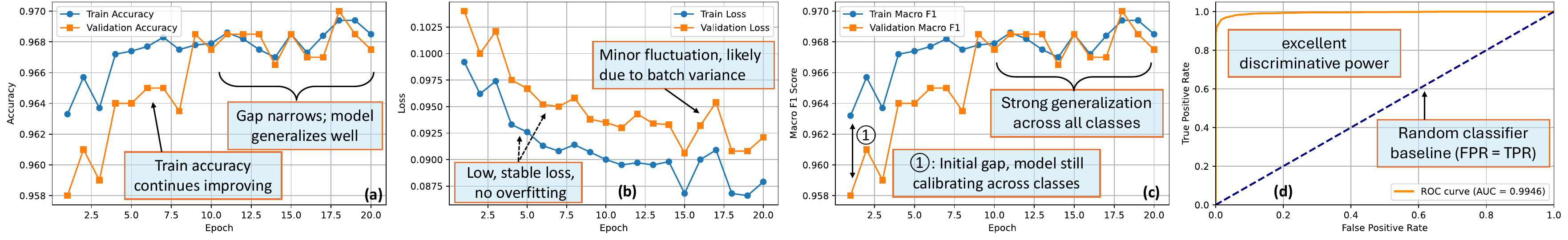}
    \caption{\footnotesize Performance evaluation of the CircuitHunt-selected model. (a) Accuracy trends over 20 epochs show early stabilization of validation accuracy and a narrowing gap with training accuracy, indicating effective generalization. (b) Loss curves exhibit a sharp initial decline in validation loss, followed by smooth convergence, reflecting stable and well-tuned optimization. (c) Macro-F1 score results demonstrate consistently high performance with minimal divergence between training and validation, highlighting the model's balanced predictive capability across classes. (d) ROC curve with an AUC of 0.9946 confirms strong discriminative power, indicating the model's ability to accurately differentiate between fraudulent and non-fraudulent cases.}
    \label{performance1}
\end{figure*}
As shown in Fig.~\ref{performance1}-a, the training and validation accuracy curves reveal a stable learning process. Validation accuracy plateaus after approximately five epochs, indicating that the model rapidly acquires discriminative features from the input data. The training accuracy continues to improve gradually, ultimately converging toward the validation curve. This narrowing gap indicates minimal overfitting and confirms that the selected circuit maintains robust generalization over time.

Fig.~\ref{performance1}-b illustrates the loss trajectories, where both training and validation loss decrease steadily over the training period. A sharp early decline in validation loss reflects effective initial learning, followed by smooth convergence from epoch 10 onward. The consistency between the two curves suggests a well-calibrated optimization process and further supports the reliability of the training setup.

The macro-F1 score, shown in Fig.~\ref{performance1}-c, provides deeper insight into the model's class-wise balance. Initially, the gap between training and validation F1-scores reflects early-stage calibration, particularly important in fraud detection where class distributions can be skewed. From epoch 10 onward, both curves stabilize above 0.965, with minimal divergence, indicating that the model performs consistently across both majority and minority classes.

The ROC curve, presented in Fig.~\ref{performance1}-d, further demonstrates the discriminative capability of the model. With an AUC of 0.9946, the model shows near-perfect ability to distinguish between fraudulent and non-fraudulent cases. The steep rise toward the top-left corner indicates high sensitivity at low false positive rates, which is critical in financial security contexts where misclassification costs are high.

Finally, class-wise test metrics are summarized in Table~\ref{tab:results}. Both the fraudulent and non-fraudulent classes exhibit high precision, recall, and accuracy, above 96.9\%, confirming the model's balanced and unbiased performance. These results affirm that the CircuitHunt pipeline effectively selects structurally valid and empirically strong quantum circuits suitable for real-world fraud detection.

\begin{table}[htbp]
    \centering
    \caption{\footnotesize Class-wise test performance of the CircuitHunt-selected model. High precision, recall, and accuracy values for both fraudulent and non-fraudulent classes confirm the model's balanced and reliable classification performance, demonstrating its effectiveness in detecting fraud without bias toward either class.}
    \label{tab:results}
    \begin{tabular}{lccc}
        \toprule
        Class & Precision & Recall & Accuracy \\
        \midrule
        Non-Fraud& 0.9704 & 0.9700 & 0.9700 \\
        Fraud & 0.9693 & 0.9696 & 0.9696 \\
        \bottomrule
    \end{tabular}
\end{table}

\subsection{Ablation Study}

To evaluate the architectural significance of the residual skip connection in our model, we conducted an ablation study by removing this component from the selected CircuitHunt architecture. The impact of this modification is systematically examined across the same metrics. 
\begin{figure*}[htpb]
    \centering
    \includegraphics[width=1\linewidth]{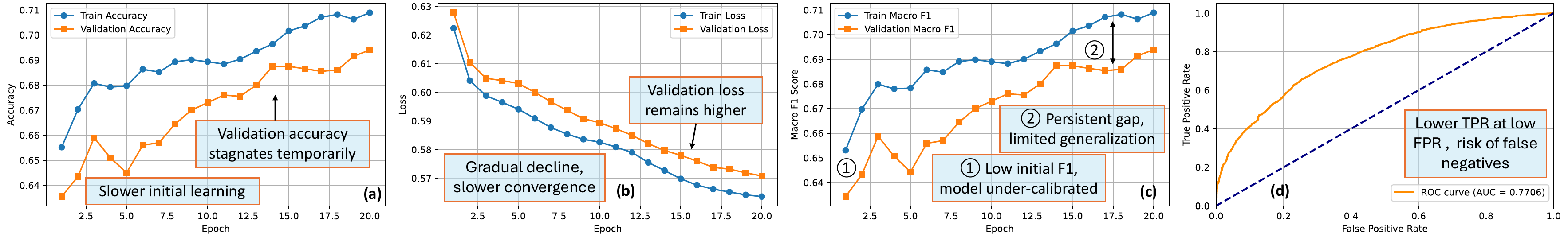}
    \caption{ \footnotesize Performance evaluation of the ablated model without the residual skip connection. (a) Training and validation accuracy curves indicate slower learning and a persistent generalization gap, underscoring the importance of the skip connection for stable convergence. (b) Loss curves show steady but suboptimal convergence, with a consistently higher validation loss compared to training. (c) Macro-F1 score trends reveal lower and less stable performance across classes, with limited generalization capacity. (d) The ROC curve shows a reduced AUC of 0.7706, highlighting the model's weakened ability to distinguish between fraudulent and legitimate instances in the absence of residual connectivity.}
    \label{ablation}
\end{figure*}
As shown in Fig.~\ref{ablation}-a, removing the skip connection results in noticeably lower training and validation accuracy. The training curve rises gradually and plateaus around 71\%, while the validation accuracy remains consistently lower, plateauing below 70\%. In contrast to the full model (see Fig.~\ref{performance1}-a), the ablated model fails to reduce the generalization gap over time, indicating reduced model expressiveness and less effective gradient propagation during optimization.

Fig.~\ref{ablation}-b further supports this observation, where both training and validation losses decline slowly and remain higher throughout training compared to the complete model (see Fig.~\ref{performance1}-b). The persistent gap between the two loss curves suggests limited learning efficiency and suboptimal convergence behavior, reinforcing the role of residual connectivity in stabilizing the training process.

The degradation in class-wise performance is also evident in the macro-F1 score trends shown in Fig.~\ref{ablation}-c. While the training F1 score increases modestly, the validation F1 lags behind, exhibiting slower growth and a wider divergence. Unlike the full model, which maintained a macro-F1 score above 0.965 (see Fig.~\ref{performance1}-c), the ablated version fails to surpass 0.70. This persistent disparity indicates that the absence of a skip connection weakens the model's ability to learn balanced representations—critical in fraud detection tasks that involve imbalanced data distributions.

The ROC curve in Fig.~\ref{ablation}-d further emphasizes the impact of the architectural change. The AUC drops to 0.7706, a substantial reduction from the 0.9946 observed with the full model (see Fig.~\ref{performance1}-d). The curve's less pronounced ascent towards the top-left corner reflects a decline in sensitivity at low false positive rates, which is particularly problematic in high-risk applications such as fraud detection. This diminished discriminative ability highlights the role of residual connectivity not only in learning stability but also in maintaining the model's decision-making precision.

\begin{table}[htbp]
    \centering
    \caption{ \footnotesize Class-wise test performance of the ablated model without the residual skip connection. Both fraudulent and non-fraudulent classes show a notable drop in precision, recall, and accuracy compared to the full model, confirming the importance of the skip connection in achieving balanced and effective fraud detection.}
    \label{tab:ablation_results}
    \begin{tabular}{lccc}
        \toprule
        Class & Precision & Recall & Accuracy \\
        \midrule
        Non-Fraud & 0.7042 & 0.7044 & 0.7044 \\
        Fraud& 0.6967 & 0.6965 & 0.6965 \\
        \bottomrule
    \end{tabular}
\end{table}
Finally, the quantitative test results in Table~\ref{tab:ablation_results} align with the observed trends. Both the fraudulent and non-fraudulent classes exhibit a drop of over 25 percentage points in precision, recall, and accuracy when compared to the full model (see Table~\ref{tab:results}), confirming the essential role played by the residual skip connection in achieving reliable and unbiased performance across classes.

Collectively, these results demonstrate that the residual skip connection is a key architectural component that significantly enhances learning dynamics, generalization ability, and class-level discrimination in quantum-enhanced fraud detection systems.

\subsection{Comparison Analysis}

To assess the empirical performance of our model, we compare it against several state-of-the-art QML architectures from recent literature. These include EstimatorQNN~\cite{innan2024financial1}, QGNN~\cite{Innan_2024}, SQNN~\cite{alami2024comparative}, and QFDNN~\cite{das2025qfdnn}, each of which has been manually designed and optimized for classification tasks relevant to financial and fraud detection applications. 
\begin{table}[htbp]
    \centering
    \caption{Comparison of existing QML architectures with our CircuitHunt-selected one.}
\begin{adjustbox}{max width=\linewidth}
\begin{tabular}{lllll}
        \toprule
        Model & Precision & Recall & F1-Score & Accuracy\\
        \midrule
                EstimatorQNN \cite{innan2024financial1} &0.89& 0.90& 0.90& 0.90 \\
       QGNN \cite{Innan_2024} & 0.96 & 0.79 & 0.86 & 0.94 \\
        SQNN \cite{alami2024comparative}  & 1.00 & 0.68 & 0.68 & 0.86 \\
        QFDNN \cite{das2025qfdnn} & 0.95  & 0.87 & 0.81 & 0.82 \\
        \textbf{CircuitHunt-Selected (Ours)} & \textbf{0.97} & \textbf{0.97} & \textbf{0.97} & \textbf{0.97} \\
        \bottomrule
    \end{tabular}
    \end{adjustbox}
    \label{tab}
\end{table}
As shown in Table~\ref{tab}, our CircuitHunt-selected model outperforms all competing architectures across all key evaluation metrics, precision, recall, F1-score, and accuracy. Notably, it achieves a balanced performance of 97\% in each metric; in contrast, other models often display trade-offs, such as SQNN achieving perfect precision but significantly lower recall, indicating poor generalization across classes.

It is important to highlight that, unlike the manually crafted models, our architecture is discovered automatically through the CircuitHunt pipeline, which searches over a space of valid quantum circuits while enforcing structural constraints and empirical validation through a macro-F1 score based selection process. This automation reduces the need for extensive design effort while also leading to superior and more consistent performance across various metrics.

These results validate the effectiveness of our approach in identifying performant quantum architectures through data-driven search rather than manual intuition. The superior balance and robustness of our model underscore the potential of automated quantum circuit design in real-world classification problems, particularly in security-critical domains such as fraud detection.

\section{Discussion}

The results collectively underscore the potential of our automated quantum circuit discovery approach, not only in achieving state-of-the-art performance but also in offering interpretability regarding design decisions. One of the most notable observations is the consistent and balanced classification performance across all major evaluation metrics. This suggests that the CircuitHunt pipeline is capable of identifying circuit structures that promote both learning stability and generalization, even under class imbalance conditions, a critical requirement in fraud detection tasks.

From a design perspective, the ablation study sheds light on the architectural value of residual skip connections in hybrid quantum-classical models. Their removal resulted in slower convergence, degraded generalization, and reduced discriminative power. This emphasizes the importance of integrating classical deep learning principles, such as residual learning, into quantum models to enhance their optimization dynamics and representational depth.

Another important takeaway is the strength of automation. While many competing quantum models rely on handcrafted circuit designs, often requiring domain knowledge and iterative tuning, our pipeline generates circuits in a principled and reproducible manner. This reduces human bias and accelerates development time while still outperforming expert-designed architectures.

Despite these promising outcomes, certain limitations remain. The current pipeline, while efficient in circuit selection, is computationally expensive during the filtering and validation stages, especially when scaling to larger architectures or datasets. Moreover, although the selected circuit generalizes well in this task, its transferability to other domains or datasets is not guaranteed and would require further investigation. Hardware noise and resource limitations may also affect real-world deployment, as the current evaluation is based on simulated environments.

In summary, the proposed approach presents a strong case for data-driven quantum circuit discovery, with clear benefits in performance, automation, and architectural insight. However, addressing computational scalability and hardware robustness will be essential steps toward broader applicability and adoption.

\section{Conclusion}

This work presents CircuitHunt, a systematic pipeline for automated quantum circuit discovery tailored for fraud detection tasks. By integrating budget-aware circuit filtering, macro-F1 score based evaluation, and a residual-enhanced hybrid architecture, the proposed approach achieves strong and balanced performance while maintaining structural feasibility and training stability. Extensive empirical evaluations, including performance benchmarking, ablation analysis, and comparative assessment, demonstrate that our automatically selected circuit outperforms manually designed counterparts and provides deeper architectural insights.

More broadly, this work paves the way for scalable and interpretable QML solutions, reducing reliance on manual design and enabling data-driven quantum architecture discovery for high-stakes applications.

\section*{Acknowledgment}
 This work was supported in part by the NYUAD Center for Quantum and Topological Systems (CQTS), funded by Tamkeen under the NYUAD Research Institute grant CG008, and the Center for Cyber Security (CCS), funded by Tamkeen under the NYUAD Research Institute Award G1104.
\bibliographystyle{IEEEtran}

\bibliography{refs}

\end{document}